\input harvmac
\input epsf

\def\~{\mathaccent "7E}


\def\figin{\epsfcheck\figin}\def\figins{\epsfcheck\figins}
\def\epsfcheck{\ifx\epsfbox\UnDeFiNeD
\message{(NO epsf.tex, FIGURES WILL BE IGNORED)}
\gdef\figin##1{\vskip2in}\gdef\figins##1{\hskip.5in}
\else\message{(FIGURES WILL BE INCLUDED)}%
\gdef\figin##1{##1}\gdef\figins##1{##1}\fi}
\def\DefWarn#1{}
\def\figinsert{\goodbreak\midinsert}
\def\ifig#1#2#3{\DefWarn#1\xdef#1{fig.~\the\figno}

\writedef{#1\leftbracket fig.\noexpand~\the\figno}%
\figinsert\figin{\centerline{#3}}\medskip\centerline{\vbox{\baselineskip12pt
\advance\hsize by -1truein\noindent\footnotefont{\bf Fig.~\the\figno:}
#2}}
\bigskip\endinsert\global\advance\figno by1}


\catcode`\@=11
\def\eqalign#1{\null\,\vcenter{\openup\jot\m@th
 \ialign{\strut\hfil$\displaystyle{##}$&
  $\displaystyle{{}##}$\hfil&&
  \qquad\hfil$\displaystyle{##}$&
  $\displaystyle{{}##}$\hfil\crcr
 #1\crcr}}\,}
\catcode`\@=12

\def\ha{{1\over 2}}

\overfullrule=0mm

\def\IR{\relax{\rm I\kern-.18em R}}
\font\cmss=cmss10 \font\cmsss=cmss10 at 7pt
\def\IZ{\relax\ifmmode\mathchoice
{\hbox{\cmss Z\kern-.4em Z}}{\hbox{\cmss Z\kern-.4em Z}}
{\lower.9pt\hbox{\cmsss Z\kern-.4em Z}}
{\lower1.2pt\hbox{\cmsss Z\kern-.4em Z}}\else{\cmss Z\kern-.4em Z}\fi}

\def\inbar{\,\vrule height1.5ex width.4pt depth0pt}
\def\IB{\relax{\rm I\kern-.18em B}}
\def\IC{\relax\hbox{$\inbar\kern-.3em{\rm C}$}}
\def\ID{\relax{\rm I\kern-.18em D}}
\def\IE{\relax{\rm I\kern-.18em E}}
\def\IF{\relax{\rm I\kern-.18em F}}
\def\IG{\relax\hbox{$\inbar\kern-.3em{\rm G}$}}
\def\IH{\relax{\rm I\kern-.18em H}}
\def\II{\relax{\rm I\kern-.18em I}}
\def\IK{\relax{\rm I\kern-.18em K}}
\def\IL{\relax{\rm I\kern-.18em L}}
\def\IM{\relax{\rm I\kern-.18em M}}
\def\IN{\relax{\rm I\kern-.18em N}}
\def\IO{\relax\hbox{$\inbar\kern-.3em{\rm O}$}}
\def\IP{\relax{\rm I\kern-.18em P}}
\def\IQ{\relax\hbox{$\inbar\kern-.3em{\rm Q}$}}
\def\IGa{\relax\hbox{${\rm I}\kern-.18em\Gamma$}}
\def\IPi{\relax\hbox{${\rm I}\kern-.18em\Pi$}}
\def\ITh{\relax\hbox{$\inbar\kern-.3em\Theta$}}
\def\IOm{\relax\hbox{$\inbar\kern-3.00pt\Omega$}}


\def\~{\tilde }
\def\^{\hat }
\def\={\bar }

\def\bbuildrel#1_#2^#3{\mathrel{\mathop{\kern 0pt#1}\limits_{#2}^{#3}}}

\catcode`\@=11
\def\displaylinesno#1{\displ@y\halign{
	\hbox to\displaywidth{$\@lign\hfil\displaystyle##\hfil$}&
	\llap{$##$}\crcr
#1\crcr}}
\catcode`\@=12

\def\myfnii{e-mail: sochen@asterix.lbl.gov }
\def\jnict{Supported by JNICT, Lisbon.}


\def\acknowledge{
This work was supported in part by the Director, Office of Energy
Research, Office of High Energy and Nuclear Physics, Division of High
Energy Physics of the U.S. Department of Energy under Contract
DE-AC03-76SF00098 and in part by the National Science Foundation  
under
grant PHY-90-21139.
}



\Title{  \hfill LBNL-39040, UCB-PTH-96/27}
{Integrable Generalized Thirring Model}
\centerline{ Korkut Bardak\c ci, Luis M. Bernardo
\footnote{$^{*}$}{\jnict} and Nir Sochen
\footnote{$^{**}$}{\myfnii} }
\vskip0.4cm
\centerline{\it Department of Physics\footnote{$^\dagger$}{\acknowledge}}
\centerline{\it University of California at Berkeley }
\centerline{\it and} 
\centerline{\it Theoretical Physics group} 
\centerline{\it Lawrence Berkeley Laboratory} 
\centerline{\it Berkeley, CA 94720, U.S.A.} 
\vskip0.4cm
\centerline{\bf Abstract }

We derive the conditions that the coupling constants of the Generalized
Thirring Model have to satisfy in order for the model to admit an infinite
number of commuting classical conserved quantities.
 Our treatment
uses the bosonized version of the model, with periodic boundary conditions
imposed on the space coordinate. Some explicit examples that satisfy these
conditions are discussed. We show that, with a different set of boundary
conditions, there exist additional conserved quantities, and we find the 
Poisson Bracket algebra  satisfied by them.

\Date{June. 96}
\def\buildrel#1\over#2{\mathrel{
     \mathop{\kern 0pt#2}\limits^{#1}}}




\newsec{Introduction}

The purpose of this paper is investigate the conditions under which the
Generalized Thirring Model becomes integrable.  The Generalized Thirring
Model is the field theory of massless fermions in two dimension,
 interacting through the
most general four fermi interaction compatible with Lorentz invariance
\ref\b{K.Bardak\c ci, Nucl.Phys. {\bf B401} (1994) 191.}. 
By integrability, we mean the existence of an infinite number of
conserved and commuting dynamical variables. In this paper, we will not 
address the question whether these are sufficient in number to make the
model truly solvable. The bosonized form of the Thirring model will be our
starting point, and our treatment from that point on will be purely classical.
It should, however, be noted that this is better than treating the original
fermionic model classically, since bosonization does capture some of the
quantum nature of the model. From the bosonized Lagrangian, we wish to
extract a Lax pair depending on a spectral parameter. For this purpose, we
first write the equations of motion in a suggestive form as flatness conditions
for two vector fields, and we demand the existence of another flat vector
field that interpolates between these two. This results in equations involving
the coupling constants which we call the first integrability condition. In
general, these equations are overdetermined and they do not have solutions
depending on a continuous parameter. In section 3, we discuss four exceptional
cases when such a solution exists. The first case is the model with maximum
internal symmetry and a single coupling constant. The second case is   a
simple generalization of the first one to a product group with two different
coupling constants. The third example is an $SU(2)$ model broken down to
$U(1)$. The last example is a model based on symmetric spaces. For each of
these examples, there exists a Lax pair depending on a spectral parameter.

Given the Lax pair, in a standard fashion, one can construct conserved
quantities in terms of a path ordered product. It is, however, necessary 
to specify boundary conditions. The simplest boundary condition is the
periodic one, with the space coordinate compactified into a circle. Taking
the trace of the path ordered product and expanding  in powers of the
spectral parameter yields an infinite number of conserved ``charges''.
In section 4, we compute the Poisson brackets of these
charges and derive the conditions so that it vanishes. This is then our second
integrability condition. In the four examples discussed earlier, the second
integrability condition is automatically satisfied, although we are unable
to prove in general that the second condition follows from the first.
Another natural boundary condition is the open one: The space interval is
from $-\infty$ to $+\infty$ and the fields vanish at $\pm \infty$. In this
case, additional integrals of motion can be constructed by considering the
matrix elements of the path ordered product, instead of just the trace. In
section 5, we compute the Poisson brackets of these additional integrals 
of motion and find that they satisfy a non-linear algebra. We also compare
this algebra to a simpler algebra derived in a somewhat similar case of
the principal chiral model \ref\SCH{J.H.Schwarz, Nucl.Phys. {\bf B447} (1995)
 137.}.
 The last section summarizes our conclusions.

\newsec{Lax Pair Formulation}

We begin this section by recalling the definition of the model, the bosonized
version of it and its equations of motion and symmetries (see \b\ and
\ref\bb{K.Bardak\c ci and L.M.Bernardo, Nucl.Phys. {\bf B450} (1995) 695.} for 
detailed analysis).
  Lightcone variables will be used throughout:
$x_{+}$ will serve as time and $x_{-}$ as space.

The parity violating generalized Thirring model is given by the action
\eqn\eIIi{S_o = \int d^{2}x(\bar{\Psi}i\gamma^{\mu}\partial_{\mu}\Psi-
(\tilde{G}^{-1})^{ab}
\; \bar{\Psi}_{R}\tilde{t}_{a}\Psi_{R}\;\bar{\Psi}_{L}\tilde{t}_{b}
\Psi_{L}),}
where $R$ and $L$ refer to the right and left chiral components of $\Psi$, 
and $\tilde{t}_{a}$ are the generators of the Lie algebra $\cal G$
 in some representation:
$$
[\tilde{t}_{a},\tilde{t}_{b}]=if_{ab}^{\ \ c}\tilde{t}_{c}.
$$
We reserve the notation $t_{a}$ for the adjoint representation.
 The coupling constant $(\tilde{G}^{-1})^{ab}$ is an invertible not 
necessarily symmetric matrix, resulting in parity violation. In one 
version of bosonization \ref\poly{A.M.Polyakov and P.B.Wiegmann, 
Phys.Lett. {\bf B131} (1983) 121, {\bf B141} (1984) 223.}\b\ref\kara
{D.Karabali, Q.H.Park and H.Schnitzer, Nucl.Phys. {\bf B323}
(1989) 572, Phys.Lett. {\bf B205} (1988) 267.}, this  gives 
\eqn\IIii{S_{o}= W(g)+W(h^{-1})
-{n\over2\pi}\int d^{2}x G_{ab}(ig^{-1}\partial_{+}g)^{a}(ih^{-1}
\partial_{-}h)^{b},}
where $X_{a}$ stands for $Tr(t_{a}X)$, $g$ and $h$ are group elements
expressed in the adjoint representation, and
\eqn\IIiii{G_{ab}={1\over2}\kappa_{ab}-{\pi\over{2n}}\tilde{G}_{ab},}
where $\kappa_{ab}$ is the Cartan-Killing metric $\kappa_{ab}=
Tr(t_{a}t_{b})$, 
$n$ is the number of fermion flavors shifted by the dual Coxeter number
and $W$ is the WZW action
\eqn\eIIiv{W(g)={n\over8\pi}\left( \int d^{2}x Tr(\partial_{\mu}g^{-1}
\partial^{\mu}g)+{2\over3}\int Tr\left( (g^{-1}dg)^{3} \right)\right).} 
The equations of motion are equivalent to conservation of two currents:
\eqn\eIIv{\partial_{+}J_{-}=\partial_{-}J_{+}=0,}
where
\eqn\eIIvi{
\eqalign{
&J_{+}  =  i{n\over4\pi}\left(-\partial_{+}h h^{-1}+
 h\tilde{t}_{a}h^{-1} (2G^T)^a_{\ b}(g^{-1}\partial_{+}g)^b\right), \cr
&J_{-}  =  i{n\over4\pi}\left(-\partial_{-}g g^{-1}+ g\tilde{t}_{a}g^{-1}
(2G)^a_{\ b}(h^{-1}\partial_{-}h)^b\right).\cr}
}
Next we notice that the model is invariant under the  transformation
$h\rightarrow u_{+}(x_{+})h$, with $J_{+}$ transforming like a gauge field.
 Using this transformation, $J_{+}$, which only depends on $x_{+}$, (see
\eIIv), can be set equal to zero. This gives us a special solution to the
equations of motion; the general solution is obtained by applying the 
inverse transformation\foot{ We thank Bogdan Morariu for clarifying
this point for us.}. For this special solution, treating $x_{+}$ as time,
we will search for an infinite set of conserved quantities, which assure
the integrability of the model. The general solution is also integrable
by virtue of the transformation introduced above. The situation is similar
to what happens in the WZW model \ref\f{L.D.Fadeev in ``Integrable Systems, 
Quantum Groups, and Quantum Field Theory'', L.A.Ibort and M.A.Rodr\'{\i}guez 
(eds.) (1993) 1, Kluwer Academic Publishers. }.
The conserved quantities are derived for this special
solution with $J_{+}=0$. 
Define now
\eqn\eIIvii{V^a_{\pm}  =  (ih^{-1}\partial_{\pm}h)^a,}
\eqn\eIIviii{W^a_{+}  = (2G^{T})^{-1a}{}_{b}V^b_{+},\;\;\; 
W^a_{-}  =  (2G)^a{}_bV^b_{-}.}
In terms of these variables, the equations of motion now read
\eqn\eIIix{\eqalign{\partial_{+}V_{-}\;-\partial_{-}V_{+}\;
-i[V_{+}\;,V_{-}\;]  &=  0, \cr
\partial_{+}W_{-}-\partial_{-}W_{+}-i[W_{+},W_{-}]  &=  0.\cr}}
These flatness conditions for the vector fields $V$ and $W$ are similar
to the zero curvature condition of integrable systems.
What is missing is a spectral parameter dependence that will provide by
power expansion an infinite number of conserved currents. We will now
derive conditions for a zero curvature with a spectral parameter dependence
(a Lax pair) to exist. This will be the first integrability condition.
 The idea is to find a one parameter family of
matrices that interpolate between the equations of motion.
It is convenient for later use to work with a special linear combination
of the $V^a_\pm$ connection:
$$
\eqalign{
M^a_+&\equiv(H_+^{\ha}(2G^T)^{-1})^a_{\ b}V_+^b,&
H_+&=1-4GG^T,\cr
M^a_-&\equiv(H_-^{\ha})^a_{\ b}V_-^b,&
H_-&=1-4G^TG.\cr}
$$
Define now an interpolating connection as follows:
\eqn\eIIx{B^a_{\pm}(x;\lambda)=N^a_{\pm b}(\lambda)M^b_{\pm}(x),}
where $N^{a}_{\pm b}$ are constants, to be determined as functions of the
spectral parameter $\lambda$.
The vector field $B$ must satisfy the zero curvature condition
\eqn\eIIxi{\partial_{+}B_{-}-\partial_{-}B_{+}-i[B_{+},B_{-}]  =  0,}
with boundary conditions
\eqn\eIIxii{\eqalign{B^a_{\pm}(x;\lambda=\lambda_0) &=V^a_{\pm}(x), \cr
B^a_{\pm}(x;\lambda=\lambda_1)&=W^a_{\pm}(x). \cr}}
In order to proceed we rewrite, after some manipulation, the equations of
motion as follows:
\eqn\eqm{
\eqalign{
\del_-M_+^a+A^a_{-pq}M_+^pM_-^q=0,\cr
\del_+M_-^a+A^a_{+pq}M_-^pM_+^q=0,\cr}
}
where
\eqn\Aabc{
A_{\pm abc}=-(H_\mp^{-\ha})_a^{\ r}(H_\mp^{-\ha})_b^{\ s}
(H_\pm^{-\ha}G_\mp)_c^{\ t}f_{rst}+
(H_\mp^{-\ha}G_\pm)_a^{\ r}(H_\mp^{-\ha}G_\pm)_b^{\ s}
(H_\pm^{-\ha})_c^{\ t}f_{rst},
}
and for convenience we defined $G_+=2G^T$ and $G_-=2G$.
Note that $A_{\pm abc}$ is antisymmetric in its first two indices.
The zero curvature reads
\eqn\xvii{N^a_{-b}\partial_{+}M^b_{-}-N^a_{+b}\partial_{-}M^b_{+}
+ f^a{}_{bc}N^b_{+p}N^c_{-q}M^p_{+}M^q_{-} = 0.}
Using equations \eqm\ and equating the coefficients of
$M^p_{+}M^q_{-}$ to zero we get the first integrability condition
\eqn\xviii{
N^a_{+b}A^b_{-pq}-N^a_{-b}A^b_{+qp}+f^a_{\ bc}N^b_{+p}N^c_{-q}=0
}
This condition gives us a Lax pair and an infinite number of resulting
 conservation laws (see the section 4). One must also show that these
conserved quantities are mutually commuting; that is,
 their Poisson brackets vanish. This will be the second 
integrability condition, derived in section 4. 

Equations \xviii\ are in general an overdetermined algebraic system with
 $(dim G)^3$ equations for $2(dim G)^2$ variables $N^{a}_{\pm b}$.
 In fact, since these equations are
nonlinear, this counting is misleading. What we have actually is a system
of polynomial equations for the variables $N^a_{\pm b}$. The locus
of the polynomials defines an algebraic variety ${\cal M}$ and the condition
of integrability is not that ${\cal M}\ne \{0\}$ (this is guaranteed since
we always have two solutions when $B$ is equal $V$ or $W$) but that 
$\dim {\cal M}\ge 1$ in order to have a spectral parameter.

Although the system is overdetermined and there are no parametric solutions 
for a generic coupling constants $G_{ab}$, there are special 
interactions for which the model is integrable. This is the subject of the 
next section.
\newsec{Solutions of the First Integrability Condition}

In this section we will construct some solutions to the integrability
condition, proving thus, the existence in those models, of an infinite
number of conserved currents. In all examples we will make use of the
diagonal ansatz.
We assume that $G$ and $N_\pm$ are diagonal matrices:
$$
\eqalign{
2G^a_{\ b}&=g_b\delta^a_{\ b},  \cr
N^a_{\pm b}&=n^\pm_b\delta^a_{\ b}, \cr}
$$
Then $A_{\pm abc}=A_{abc}$,
\eqn\Aabcdiag{
A_{abc}={{g_ag_b-g_c}\over{(1-g^2_a)^\ha(1-g^2_b)^\ha(1-g^2_c)^\ha}}f_{abc},
}
and the integrability condition \xviii\ reduces to
$$
\eqalign{
n^+_aA_{abc}-n^-_aA_{acb}+f_{abc}n^+_bn^-_c =0, \cr}
$$
or, using \Aabcdiag\ ,
\eqn\nplusa{
\big( n^+_a{{g_ag_b-g_c}\over{(1-g^2_a)^\ha(1-g^2_b)^\ha(1-g^2_c)^\ha}}
      +n^-_a{{g_ag_c-g_b}\over{(1-g^2_a)^\ha(1-g^2_b)^\ha(1-g^2_c)^\ha}}
      +n^+_bn^-_c \big)f_{abc}=0. }
Note that there is no implied sum here.

$\underline{\hbox{Example 1}}$: $2G=g{\bf 1}$. (The Symmetric Case)

In this case, $N_\pm=n^\pm{\bf 1}$, and we have one equation for two
variables,
\eqn\exampone{
n^+{{g^2-g}\over{(1-g^2)^{3\over2}}}+n^-{{g^2-g}\over{(1-g^2)^{3\over2}}}
    +n^+n^-=0,
}
with a one parameter family of solutions.

$\underline{\hbox{Example 2}}$: $2G=g_1{\bf 1}\otimes g_2{\bf 1}$

This just gives two copies of the above equation \exampone\ .

$\underline{\hbox{Example 3}}$: $SU(2)$ with $U(1)$ symmetry.
 $2G^a_{\ b}=g_b\delta^a_{\ b}$ with $g_1=g_2\ne g_3$.

This suggests the  ansatz, $N^a_{\pm b}=n^\pm_b\delta^a_{\ b}$ with
$n^\pm_1=n^\pm_2\ne n^\pm_3$. Then there are three equations for four
variables,
\eqn\exampthree{
\eqalign{
n^+_1(g^2_1-g_3)+n^-_1(g_1g_3-g_1)+(1-g^2_1)(1-g^2_3)^\ha n^+_1n^-_3 &=0, \cr
n^+_3(g_3g_1-g_1)+n^-_3(g_3g_1-g_1)+(1-g^2_1)(1-g^2_3)^\ha n^+_1n^-_1 &=0, \cr
n^+_1(g_1g_3-g_1)+n^-_1(g^2_1-g_3)+(1-g^2_1)(1-g^2_3)^\ha n^+_3n^-_1 &=0. \cr}
}
The one parameter family of solutions is given by,
\eqn\solutions{
\eqalign{
n^+_1 &=\left({g_1\over{\lambda (1+g_3)(1-g^2_1)^2}}\big( 2(g_3-g^2_1)
   + {1\over \lambda}g_1(1-g_3) +g_1(1-g_3)\lambda \big) \right)^\ha , \cr
n^-_1 &=\left({{g_1\lambda}\over{(1+g_3)(1-g^2_1)^2}}\big( 2(g_3-g^2_1)
   + {1\over \lambda}g_1(1-g_3) +g_1(1-g_3)\lambda \big) \right)^\ha ,\cr
n^+_3 &={1\over{(1-g^2_3)^\ha (1-g^2_1)}}\big( (g_3-g^2_1)+{1\over \lambda}
  g_1(1-g_3)\big),\cr
n^-_3 &={1\over{(1-g^2_3)^\ha (1-g^2_1)}}\big( (g_3-g^2_1)+\lambda
  g_1(1-g_3)\big), \cr}
}
where $\lambda={n^-_1\over n^+_1}$ is the free parameter with range 
${1 \over g_1} \le \lambda \le g_1$. Note that the case where we also have
$g_2 \ne g_1$ gives six equations for six variables.

$\underline{\hbox{Example 4}}$: Symmetric spaces

Let $F$ be a simple group with a subgroup $H$. Then the Lie algebra ${\cal{F}}$
can be decomposed into the Lie algebra ${\cal{H}}$ and its orthogonal
complement ${\cal{K}}$, which generates the coset $F/H$. This coset space is
a symmetric space if $[{\cal{K}},{\cal{K}}]\subset {\cal{H}}$. In what follows
we will label the generators of $H$ with greek indices (i.e. $\lambda^\alpha$),
and the generators of $F/H$ with latin indices (i.e. $\lambda^a$), and when we
don't want to specify between them, we'll use dotted latin indices (i.e.
$\lambda^{\dot a}$).

To have a coset (symmetric) space in the present model we choose
$$
\eqalign{
2G^{\dot a}_{\ {\dot b}}&=g_{\dot b}\delta^{\dot a}_{\ {\dot b}},  \cr}
$$
with $g_a=g$ and $g_\alpha=1$. This assures that the currents in ${\cal{H}}$
are set to zero, resulting in a coset model.
We cannot use \xviii\ because $A_{\pm abc}$ is not defined in this case 
($H_\pm$ is not invertible), but instead, we have to use \eIIix\ and \eIIxi\ :
\eqn\symm{
\eqalign{
\partial_+V^{\dot a}_--\partial_-V^{\dot a}_++f^{\dot a}_{\ {\dot b}
    {\dot c}}V^{\dot b}_+V^{\dot c}_- &=  0, \cr
\partial_+W^{\dot a}_--\partial_-W^{\dot a}_++f^{\dot a}_{\ {\dot b}
    {\dot c}}W^{\dot b}_+W^{\dot c}_- &=  0, \cr
\partial_+B^{\dot a}_--\partial_-B^{\dot a}_++f^{\dot a}_{\ {\dot b}
    {\dot c}}B^{\dot b}_+B^{\dot c}_- &=  0. \cr}}
The choice of $G$ leads to the ansatz:
$$
\eqalign{
B^{\dot a}_{\pm}(x;\lambda)&=r^{\pm}_{\dot a}(\lambda)
V^{\dot a}_{\pm}(x),  \cr}
$$
with $r^{\pm}_a(\lambda)=r^{\pm}(\lambda)$ and $r^{\pm}_{\alpha}
(\lambda)=1$. Then, we can rewrite \symm\ as
\eqn\expsymm{
\eqalign{
(1-g^2_{\dot a})\partial_+V_{- \dot a}+f_{\dot a \dot b \dot c}
(1-{{g_{\dot a}g_{\dot c}}\over{g_{\dot b}}})V_{+\dot b}V_{-\dot c}&=0, \cr
(1-{1\over{g^2_{\dot a}}})\partial_-V_{+ \dot a}+f_{\dot a \dot b \dot c}
(1-{{g_{\dot c}}\over{g_{\dot a}g_{\dot b}}})V_{+\dot b}V_{-\dot c}&=0, \cr
r^-_{\dot a}\partial_+V_{-\dot a}-r^+_{\dot a}\partial_-V_{+\dot a}
+f_{\dot a \dot b \dot c}r^+_{\dot b}r^-_{\dot c}V_{+\dot b}
V_{-\dot c}&=0. \cr}}
Note that $\dot a$ is fixed (no implied sum). Now if $\dot a $ is in 
${\cal{H}}$, then the first two equations in \symm\ are the same, and from the
third one we get the condition $r^+(\lambda)r^-(\lambda)=1$. The fact that the
first two equations become the same is a sign of gauge invariance. To fix
a gauge we choose $V_{-\alpha}=0$. Then in this gauge, if 
$\dot a$ is in ${\cal{K}}$, the first two equations 
in \expsymm\ solve the third one without further conditions. 
Hence, for symmetric spaces, we get one equation for two variables
\eqn\oeftv{r^+(\lambda)r^-(\lambda)=1.}


\newsec{The Poisson Bracket -- Periodic Boundary Condition}

Having a Lax pair at hand, we can construct conserved quantities in the 
time variable $x_{+}$, if we also impose periodicity in the space coordinate
$x_{-}$. We first define the following quantity:
\eqn\Pexp{
{\cal{U}}(x_{+},x_{-};\lambda)=Pe^{-i\int^{x_{-}}_{x_{-}}B_{-}(x_{+},x'_{-}
;\lambda)dx'_{-}},
}
and  take $B_\pm$ periodic in $x_-$ with period $2\pi$. The
integral goes between $x_-$ and $x_-+2\pi\sim x_-$. This quantity satisfies
the equation
\eqn\delminus{
\del_- {\cal{U}}-i[B_-,{\cal{U}}]=0,
}
which can be taken as the definition
\foot{This is the same as defining the
path ordered exponential with the ``upper'' limit fixed, i.e.,
$ Pe^{\int^b_a dx A(x)} = 1 + \int^b_a dx A(x) + \int^b_a dx \int^b_x dx'
A(x) A(x') + \cdots $. Note that this is the opposite of the usual definition
where the ``lower'' limit is fixed.}
of path ordering in \Pexp.
 More importantly, the trace of this matrix,
$U(\lambda)=\Tr\ {\cal{U}}(x_+,x_-;\lambda)$, is conserved:
\eqn\delplus{
\eqalign{
\del_+ U&=i\int_{x_-}^{x_-}\Tr(\del_+B_-{\cal{U}})dx'_-\cr
(\hbox{Bianchi id.})&=i\int_{x_-}^{x_-}\Tr\big((\del_-B_++i[B_+,B_-])
\,{\cal{U}}\big)
dx'_-\cr
(\hbox{integration by parts})&=-i\int_{x_-}^{x_-}
\Tr\big(B_+(\del_-{\cal{U}}-i[B_-,{\cal{U}}])\big)dx'_-\cr
(\hbox{by \delminus})&= 0.\cr}
}
We see that $U$ is a conserved quantity and upon expanding it in powers of
$\lambda$ we get infinite number of conserved currents. Another possible
way of of increasing the list of conserved quantities is to choose $B$ to 
live in different representations of the Lie algebra.
Since $U$ does not 
depend on $x_+$, we drop its dependence from $U$ as well as the subindex
$-$ from $x_-$. It is convenient to think on the $+$ direction as ``time"
and the $-$ direction as ``space". The next step is to find the algebra of the
conserved currents. In particular if they all commute then we have infinite 
number of conserved quantities in involution, which is the trademark of 
integrability. We want to calculate the Poisson brackets of 
$U(\lambda)$ with $U(\mu)$:
\eqn\expcomi{
\eqalign{
\{U(\lambda)\ ,\ U(\mu)\}&=\int dxdy
{\delta U(\lambda)\over \delta B_-^a(x;\lambda)}
{\delta U(\mu)\over \delta B_-^b(y;\mu)}
\{ B_-^a(x;\lambda), B_-^b(y;\mu)\} \cr
&=-\int dxdy U_a(x;\lambda)U_b(y;\mu)
\{ B_-^a(x;\lambda), B_-^b(y;\mu)\} \cr
&=-\int dxdy U_a(x;\lambda)U_b(y;\mu)N_{-c}^a(\lambda)N_{-d}^b(\mu)
\{ M_-^c(x), M_-^d(y)\}. \cr}
}
The Poisson brackets for the $M$'s where calculated in \b\ :
\eqn\Barps{
\eqalign{
\{M_-^c(x),M_-^d(y)\}=&-{4\pi\over n}\kappa^{cd}\delta'(x-y)+
{4\pi\over n}F^{cd}_{+ \ k}M_-^k(x)\delta(x-y) \cr
&+{2\pi\over n}\epsilon(x-y)E^{cd}_{\ \ kl}(x,y)M_-^k(x)M_-^l(y),\cr}}
where
\eqn\edefs{
\eqalign{
&E^{cd}_{\ \ kl}(x,y)=A_{+k}^{\ c \ r}A_{+\,l}^{\ d \ s}
\big(Pe^{\int_x^y A_{-a}M_-^a(x')dx'}\big)_{rs}\quad,\quad
(A_{-c})_{ab}=A_{-abc},\cr
&F_{+ abc}=
(H_-^{-\ha})_a^{\ r}(H_-^{-\ha})_b^{\ s}
(H_-^{-\ha})_c^{\ t}f_{rst}-
8(H_-^{-\ha}G^T)_a^{\ r}(H_-^{-\ha}G^T)_b^{\ s}
(H_-^{-\ha}G^T)_c^{\ t}f_{rst}\hfil .\cr}
}
Notice that the Poisson brackets include a non local term! It is clear that
the only way to get a local algebra for the Poisson brackets for our conserved
fields is that the coefficient of $\epsilon(x-y)$ is a derivative such that
by integration by parts the derivative acts on $\epsilon$ to give a $\delta$
function. For a generic coupling this can not happen, but exactly for the
models that satisfy the integrability condition this is true! 
\eqn\ansatz{
\eqalign{
\{U(\lambda)\ &,\ U(\mu)\}_{non-local}=\cr
=&{2\pi\over n}
\int dxdy U_a(x;\lambda)U_b(y;\mu)N_{-c}^a(\lambda)N_{-d}^b(\mu)
\epsilon(x-y)E^{cd}_{\ \ kl}(x,y)M_-^k(x)M_-^l(y)\cr
=&{2\pi\over n}\int dxU_a(x;\lambda)N_{-c}^a(\lambda)A_{+k}^{\ c \ r}
M^k_-(x)\cr
&\int dy\epsilon(x-y)U_b(y;\mu)(N^b_{-d}(\mu)A_{+\,l}^{\ d \ s})
\big(Pe^{\int_x^yA_{-k}M_-^k(x')dx'}\big)_{rs}M_-^l(y)\cr
=&{2\pi\over n}\int dxU_a(x;\lambda)N_{-c}^a(\lambda)A_{+k}
^{\ c \ r}M^k_-(x)\cr
&\int dy\epsilon(x-y)U_b(y;\mu)(N^b_{+d}(\mu)A^{\ ds}_{-\ \, l}+
f^b_{\ tu}N_+^{ts}(\mu)N_{-l}^u(\mu)) \cr
&\times\big(Pe^{\int_x^yA_{-k}M_-^k(x')dx'}\big)_{rs}M_-^l(y),\cr}
}
where the last step follows from the integrability condition, \xviii\ .
Using the following identities
\eqn\iden{
\eqalign{
&\del_yU_t(y;\mu)=-f_{\ tu}^bB^u_-(y;\mu)U_b(y;\mu)=
-f_{\ tu}^bN^u_{-l}(\mu)M_-^l(y)U_b(y;\mu),\cr
&\del_y\big(Pe^{\int_x^yA_{-k}M_-^k(x')dx'}\big)_r^{\ d}=
-A_{-\ \,l}^{\ ds}M_-^l(y)\big(Pe^{\int_x^yA_{-k}M_-^k(x')dx'}\big)_{rs},\cr}
}
we can write now the integrand of the $y$ integral as a total derivative
\eqn\rside{
\eqalign{
&U_b(y;\mu)\left(N^b_{+d}(\mu)A^{\ ds}_{-\ \,l}+
f^b_{\ tu}N_+^{ts}(\mu)N_{-l}^u(\mu)\right)
\big(Pe^{\int_x^yA_{-k}M_-^k(x')dx'}\big)_{rs}M_-^l(y)=\cr
&-\del_y\Big(U_b(y;\mu)N^b_{+d}(\mu)\big(Pe^{\int_x^yA_{-k}M^kdx'}
\big)_r^{\ d}\Big).\cr}
}
The contribution of the non-local piece is therefore
\eqn\nonloc{
\eqalign{
\{U(\lambda)\ ,&\ U(\mu)\}_{non-local}
={4\pi\over n}
\int dx U_a(x;\lambda)U_b(x;\mu)N^a_{-c}(\lambda)N^b_{-d}(\mu)
A_{+\,l}^{\ c \ d}M_-^l(x)\cr
&={2\pi\over n}\int dx U_a(x;\lambda)U_b(x;\mu)\left(
N_{-c}^a(\lambda)N_{+d}^b(\mu)A_{+  \,l}^{\ c \ d}
-N_{+c}^a(\lambda)N_{-d}^b(\mu)A_{+ \, l}^{\ d \ c}\right)M_-^l(x),\cr}
}
where in the second step we (anti)symmetrized the right hand side so that the
change in  sign under $\lambda \leftrightarrow \mu$ is manifest.
The contributions from the local terms are easily evaluated:
\eqn\eqdelt{
\{U(\lambda)\ ,\ U(\mu)\}_{\delta}
=-{4\pi\over n}
\int dx U_a(x;\lambda)U_b(x;\mu)N_{-c}^a(\lambda)N_{+d}^b(\mu)
F_{+ \ l}^{cd}M_-^l(x),
}
and
\eqn\eqdeltp{
\eqalign{
\{U(\lambda)\ & ,\ U(\mu)\}_{\delta'}
={2\pi\over n}
\int dx U_a(x;\lambda)U_b(x;\mu)\cr
&\left(f^a_{\ rs}N_{-l}^s(\lambda)
N_{-c}^r(\lambda)N_{-d}^b(\mu)\kappa^{cd}
- f^b_{\ rs}N_{-l}^s(\mu)
N_{-c}^a(\lambda)N_{-d}^r(\mu)\kappa^{cd}\right)M_-^l(x).\cr}
}
Putting everything together we get
\eqn\together{
\{U(\lambda)\  ,\ U(\mu)\}=\int dx U_a(x;\lambda)U_b(x;\mu)J^{ab}_{\ \ \,l}
(\lambda,\mu)M_-^l(x),
}
where
\eqn\Jabc{
\eqalign{
J^{ab}_{\ \ \,l}(\lambda,\mu)=&
N_{-c}^a(\lambda)N_{+d}^b(\mu)A_{+ \, l}^{\ c \ d}-
N_{+c}^a(\lambda)N_{-d}^b(\mu)A_{+ \, l}^{\ d \ c}-
2N_{-c}^a(\lambda)N_{-d}^b(\mu)F_{+ \ l}^{cd}\cr
&+ f^a_{\ rs}N_{-l}^s(\lambda)N_{-c}^r(\lambda)N_{-d}^b(\mu)\kappa^{cd}
-f^b_{\ rs}N_{-l}^s(\mu)N_{-c}^a(\lambda)N_{-d}^r(\mu)\kappa^{cd}.\cr}
}
The right hand side of equation \together\ is zero, if the integrand is a
total derivative. This motivates us to try the ansatz
\eqn\eqansatz{
\del_x\big(U_a(x;\lambda)U_b(x;\mu)C^{ab}(\lambda,\mu)\big)=
U_{a}(x;\lambda)U_{b}(x;\mu)J^{ab}_{\ \ \,l}(\lambda,\mu)M^{l}_{-}(x).
}
The left hand side of this equation can be computed using the identities
of \iden\ . This result in the second integrability condition
\eqn\sdintcon{
J^{ab}_{\ \ \,l}(\lambda,\mu)=
f^a_{\ cd}N_{-l}^d(\lambda)C^{cb}(\lambda,\mu)+
f^b_{\ cd}N_{-l}^d(\mu)C^{ac}(\lambda,\mu).
}

In the diagonal ansatz, $C$ has the form
$$
\eqalign{
C^{ab}(\lambda,\mu)&=C^a(\lambda,\mu)\kappa^{ab}.  \cr}
$$
For models for which a $C^{ab}$ that satisfies this equation can be found,
$\{U(\lambda),U(\mu)\}=0$, and the conserved quantities are in involution.
We will now check  the models that satisfy the first integrability condition
against this second integrability condition. In all examples we have, the
result is the same: The second integrability condition is automatically
satisfied. However, we do not know whether in general the second condition
follows from the first one.

 In example 1, the symmetric
model, we take $ C^{a}(\lambda, \mu)= C(\lambda,\mu)$, and
then  we have one equation for one unknown, $C(\lambda,\mu)$. The condition
\sdintcon\ is satisfied, and the conserved quantities are commutative. The
generalization to example 2 is trivial. In case of example 3,
 $SU(2)$ with $C^{ab}(\lambda,\mu)=C^a(\lambda,\mu)\kappa^{ab}$ and with
$C^1(\lambda,\mu)=C^2(\lambda,\mu)\ne C^3(\lambda,\mu)$,
 initially we have three equations for two unknowns, but using the fact
that $C^a(\lambda,\mu)=-C^a(\mu,\lambda)$ we see that there are only two
equations. So, in this case also, equation \sdintcon\ can be solved, and
the model is integrable. We have not checked example 4 in detail, but we
suspect that the result is the same.

\newsec{The Poisson Bracket -- Open Boundary Condition}

In this section we will show that the structure of the model is richer with
open boundary conditions. We consider the open interval from $-\infty$ to
$+\infty$, with fields vanishing at $\pm\infty$.
With these conditions, it is possible to construct more integrals of motion
than with the periodic one. These integrals do not commute in general and it is
the aim of this section to calculate the algebra they generate.
A subset of these integrals are the integrals we saw in the periodic
case, and they  commute, insuring, thus, the integrability of the model.

The calculation of the Poisson brackets is the same as the periodic case, the
only difference will come from the boundary. Defining
\eqn\Pexpob{
{\cal{U}}_\lambda(x,y)= Pe^{-i\int_{x}^{y}B_-(x';\lambda)dx'},
}
then, any matrix element $U^{ab}(\lambda)={\cal{U}}^{ab}_\lambda
(-\infty,\infty)$, and not just
the trace, is conserved, provided that $B_{+}(\pm\infty,\lambda)=0$.
This follows from manipulations similar to \delplus, noticing that the
boundary terms coming from $\pm\infty$ can be dropped.
Of course, $B_{-}$ also must vanish sufficiently fast at $\pm \infty$ for
the integral to exist. For the sake of simplicity, we will also take $B$ in
the adjoint representation.

Define now
\eqn\eqconobi{
U^{ab}_{\ \ c}(x;\lambda)=\big( {\cal{U}}_\lambda(-\infty,x)t_{c}
{\cal{U}}_\lambda(x,\infty)\big)^{ab},
}
then
\eqn\expcomiob{
\eqalign{
\{U^{aa'}(\lambda)\ ,\ U^{bb'}(\mu)\}&=\int dxdy
{\delta U^{aa'}(\lambda)\over \delta B_-^c(x;\lambda)}
{\delta U^{bb'}(\mu)\over \delta B_-^d(y;\mu)}\{B^c_-(x;\lambda),
B^d_-(y;\mu)\}\cr
&=-\int dxdy U^{aa'}_{\ \ c}(x;\lambda)U^{bb'}_{\ \ d}(y;\mu)
N_{-k}^c(\lambda)N_{-l}^d(\mu)
\{ M_-^k(x), M_-^l(y)\}. \cr}
}
Being careful to take into account the boundary contribution from the 
non-local term in the Poisson brackets of the $M$'s, we get
\eqn\together{
\eqalign{
\{U^{aa'}(\lambda)\  ,\ U^{bb'}(\mu)\}=&{2\pi\over n}\int dx\; U^{aa'}_{\ \ c}
(x;\lambda)U^{bb'}_{\ \ d}(x;\mu)
J^{cd}_{\ \ e}(\lambda,\mu)M_-^e(x)\cr
&+{2\pi\over n}\big(U(\lambda)t_c\big)^{aa'}\big(t_dU(\mu)\big)^{bb'}
U^{lk}(0)N_{-k}^c(\lambda)N_{-l}^d(\mu)\cr
&-{2\pi\over n}\big(t_cU(\lambda)\big)^{aa'}\big(U(\mu)t_d\big)^{bb'}
U^{kl}(0)N_{-k}^c(\lambda)N_{-l}^d(\mu),\cr} }
where $J^{ab}_{\ \ c}$ is defined in \Jabc , and   $U^{ab}(0)= U^{ab}
(\lambda=\lambda_0)$. If $J^{ab}_{\ \ c}$ also satisfies
the second integrability condition \sdintcon\ then the integrand can be written
as a derivative. The only contributions come from the boundary and the final
result is
\eqn\finalob{
\eqalign{ \{U^{aa'}(\lambda)\  ,\ U^{bb'}(\mu)\}={2\pi\over n}\Big(
&\big(U(\lambda)t_c\big)^{aa'}\big(U(\mu)t_d\big)^{bb'}
C^{cd}(\lambda,\mu)\cr
-&\big(t_cU(\lambda)\big)^{aa'}\big(t_dU(\mu)\big)^{bb'}C^{cd}(\lambda,\mu)\cr
+&\big(U(\lambda)t_c\big)^{aa'}\big(t_dU(\mu)\big)^{bb'}
U^{lk}(0)N_{-k}^c(\lambda)N_{-l}^d(\mu)\cr
-&\big(t_cU(\lambda)\big)^{aa'}\big(U(\mu)t_d\big)^{bb'}
U^{kl}(0)N_{-k}^c(\lambda)N_{-l}^d(\mu)\Big).\cr}
}
This algebra, in the symmetric case, can be compared with the algebra obtained 
without bosonizing the fermions \ref\devega{H.J.de Vega, 
H.Eichenherr and  J.M.Maillet, Phys.Letters {\bf B132} (1983) 337 ,
Comm.Math.Phys. {\bf 92} (1984) 507, Nucl.Phys. {\bf B240} (1984) 377.}. 
Both the 
appearance of cubic terms and the more complicated spectral dependence of
the ``classical'' $r$-matrix $C^{ab}(\lambda,\mu)$ are quantum effects 
due to the process of bosonization.

Equation \finalob\ exhibits a closed algebra for the matrix elements 
$U^{ab}(\lambda)$; $N$'s and $C$'s are the corresponding ``structure 
constants'' that depend on the 
particular model under consideration. The algebra is clearly
non-linear, with quadratic and cubic terms in $U$ appearing on the right hand
side. We have not succeeded in identifying it with any well-known algebra,
although it may possibly bear some relation to the W algebras 
\ref\W{A.B.Zamolodchikov, Theor.Math.Phys. {\bf 63} (1985) 1205.}.
It is also the generalization of the algebra L\"{u}scher and Pohlmeyer 
\ref\LP{M.L\"{u}scher and K.Pohlmeyer, Nucl.Phys.{\bf B137} (1978) 46.} found
for the special case of the fundamental  representation of $SU(2)$.
Finally we would like to comment on the relation of this algebra to the
 affine algebra of currents  found in the principal
chiral model \SCH. In the symmetric case (Example 1
of section 3), the Thirring model bears a great resemblance to the 
principal chiral model; for example, the equations of motion \eIIix\ in
this case reduce to
\eqn\motion{
\eqalign{
g\del_+ V_- + \del_- V_+ &= 0, \cr
\del_+ V_- -\del_- V_+ -i[V_+,V_-] &= 0.\cr}
}
For $g=1$, these are identical to the equations of motion of the principal
chiral model. Not surprisingly, the Lax pair and the conserved quantities
are in one to one correspondence, with obvious modifications in case 
$g\neq 1$. However, the algebra \finalob\ is quite different from the affine
algebra found in \SCH. This is because in \SCH, the Poisson brackets
of the conserved quantities were derived from the transformations they
generate on the field variables, whereas we have used the standard Poisson
 structure given by the Lagrangian. The two Poisson structures differ in 
this case, as well as in the principal chiral model
\ref\nir{N.Sochen, ``Integrable Generalized Principal Chiral Model'',
LBL-38034, UCB-PTH-96/04.}.

\newsec{Conclusion}

In this paper we derived very general conditions (equations \xviii\ and 
\sdintcon\ and \Jabc) that the coupling constant of the Generalized
Thirring Model should satisfy to be integrable, in the sense of having an
infinite number of conserved quantities. Some solutions were
found for special cases. The algebra generated by the conserved currents
was calculated for both periodic and open boundary conditions. In the
periodic case the currents are in involution while in the open case we
found a non linear algebra which has an Abelian subalgebra of conserved
currents in involution.

The directions for further research are numerous. One
may, for example,  add fermions  to the bosonized theory and study 
supersymmetric models. Or, one can
try to generalize to a space dependent metric \ref\NS{N.Sochen, ``Integrable 
Generalized Sigma Models on a Group Manifold'', in preparation.}. There is 
of course the problem of finding new solutions to the overdetermined
algebraic systems eqs. \xviii, \sdintcon\ and \Jabc.
It is important to understand the non linear algebra between the conserved
quantities that emerges in
the open boundary case..
We would also like to quantize these models and to study the quantum 
algebraic structure.
As a first check we would like to see if the relation between the coupling
constants that ensures integrability holds along the
renormalization group flow. For the solutions we have studied in this paper,
this result follows rather trivially from symmetry considerations.
Finally there are the questions of the 
relation to 
irrational conformal field theories
(see \ref\H{M.B.Halpern, E.Kiritsis, N.A.Obers and K.Clubok,
Phys.Rep. {\bf 265} (1996) 1,
and hep-th/9501144} and references therein), and to integrable perturbations
of conformal field theories.  We hope to address some of these
questions in the future.

\listrefs
\bye